\documentclass[
 aip,
 amsmath,amssymb,
 reprint,
 floatfix,
]{revtex4-1}
\usepackage{graphicx}
\usepackage{dcolumn}
\usepackage{bm}
\usepackage[utf8]{inputenc}
\usepackage[T1]{fontenc}
\usepackage{mathptmx}
\usepackage{etoolbox}
\usepackage{physics}
\makeatletter
\def\@email#1#2{
 \endgroup
 \patchcmd{\titleblock@produce}
  {\frontmatter@RRAPformat}
  {\frontmatter@RRAPformat{\produce@RRAP{*#1\href{mailto:#2}{#2}}}\frontmatter@RRAPformat}
  {}{}
}
\usepackage{multirow}
\usepackage[]{color}
\usepackage{soul}
\usepackage{cancel,siunitx}

\makeatother
\begin{document}

\preprint{AIP/123-QED}

\title{On the characteristic features of ionization in QED environments}
\author{Rosario R. Riso}
\author{Tor S. Haugland}%
\affiliation{ 
Department of Chemistry, Norwegian University of Science and Technology, 7491 Trondheim, Norway%\\This line break forced with \textbackslash\textbackslash
}%

\author{Enrico Ronca}
\affiliation{%
Istituto per i Processi Chimico Fisici del CNR (IPCF-CNR), Via G.Moruzzi, 1, 56124, Pisa, Italy%\\This line break forced% with \\
}%

\author{Henrik Koch*}
\affiliation{%
Scuola Normale Superiore, Piazza dei Cavalieri 7, 56126 Pisa, Italy%\\This line break forced% with \\
}%https://www.overleaf.com/project/619510c59c313c6893c80501
 \email{henrik.koch@sns.it}

%\date{\today}

\begin{abstract}
The ionization of molecular systems is important in many chemical processes such as electron transfer and hot electron injection. Strong coupling between molecules and quantized fields (e.g. inside optical cavities) represents a new promising way to modify molecular properties in a non-invasive way. Recently, strong light-matter coupling has shown the potential to significantly improve the rates of hot electron driven processes, for instance in water splitting. 
In this paper, we demonstrate that inside an optical cavity 
the residual interaction between an outgoing free electron and the vacuum field is significant. We further show that, since the quantized field is also interacting with the ionized molecule, the free electron and the molecular system are correlated. We develop a theoretical framework to account for the field induced correlation and show that the interaction between the free electron and the field free electron-field interaction has sizeable effects on the ionization potential of typical organic molecules. 
\end{abstract}

\maketitle

\section{Introduction}\label{sec:introduction}

Strong coupling between molecules and quantized fields has proven to be a very effective way to engineer molecular properties.\cite{Polaritons_Scholes,Quantum_Wilson,Cavity_Hagenmuller,Cavity_Wang} Possible applications range from the control of photochemical processes \cite{kowalewski2017manipulating,Modyfying_Hutchison,Investigating_Mandal}  to the modification of molecular reactivity.\cite{Novel_Bennett,Cavity_Galego, Coherent_Shalabney,Groundstate_Thomas,Cavity_Lather,Tilting_Thomas} 
The easiest way to achieve strong coupling is through optical cavities, devices composed of mirrors confining the electromagnetic radiation in a reduced volume.\cite{Laser_Siegman,du2019remote,Quantum_Fitzgerald,chikkaraddy2016single} Inside an optical cavity the photonic vacuum couples to the molecular system creating mixed light-matter states called polaritons.\cite{Light_Wang,ruggenthaler2018quantum,herrera2020molecular} 
Since the properties of polaritonic states can be tuned\cite{Cavity_Herrera,ebbesen2016hybrid} by changing the field inside the resonator, polaritonic chemistry promises to be a non-invasive methodology to modify molecular properties on demand.\cite{Polariton_Yuen,latini2019cavity,fregoni2018manipulating,ribeiro2018polariton} The theoretical comprehension of phenomena in the strong coupling regime is still in its infancy and \textit{ab initio} approaches to study strong light-matter interaction have only been developed in recent years.\cite{Coupled_Haugland,Atoms_Flick,Quantum_Ruggenthaler,Intermolecular_Haugland,Polarized_Mandal,Cavity_Ashida,pavosevic2022cavity} 
Under strong coupling conditions the electromagnetic field is a crucial part of the system. For this reason, the field must be treated on the same footing 
as the electrons, that is, following quantum electrodynamics (QED) prescriptions.\cite{loudon2000quantum} A widely used strategy to tackle problems in the strong coupling regime is to take inspiration from standard quantum chemistry theories. Indeed, many concepts can be generalized, in a relatively simple way, to QED environments (i.e. QED Hartree Fock,\cite{Coupled_Haugland} QED coupled cluster,\cite{Intermolecular_Haugland,Coupled_Haugland,pavosevic2022cavity,pavosevic2021polaritonic} QED density functional theory \cite{Atoms_Flick,Quantum_Ruggenthaler,Ab_Flick,Schafere2110464118}). However, instances where this generalization procedure is nontrivial can also arise.

Ionization is a key process in chemistry, as electron removal is used to follow the advancement of chemical reactions, or to characterize molecular systems in spectroscopic techniques such as the X-ray photoelectron spectroscopy \cite{ghodselahi2008xps} (XPS). \cite{agarwal2013x,gulino2013structural,moitra2022multi,lewis2011dissociation} Additionally, molecular ionization can be used to initiate and promote new reactive pathways.\cite{liekhus2015ultrafast,kim2012x}
In some recent papers, DePrince,\cite{deprince2021cavity} Liebenthal \textit{et al.} \cite{Equation_Liebenthal} and Pavovsevic \textit{et al.} \cite{pavosevic2021polaritonic,pavosevic2022cavity} demonstrated that ionization potentials and electronic affinities change inside optical cavities. This is of particular relevance in the context of cavity QED since the pioneering work by Shi \textit{et al.}\cite{shi2018enhanced} has shown that the production of hot electrons under strong coupling conditions is a viable way to improve water splitting processes.
While detailed theoretical descriptions of ionization processes outside optical cavities are available,\cite{saitow2019accurate,melania2007dyson,landau2010frozen,coriani2015communication,ronca2017time,krylov2008equation} much work is still needed for QED environments as the free electron will still interact with the electromagnetic field.

In this work we present a new definition of ionization potentials for strongly coupled systems. In particular, we demonstrate that a quantized field induces sizeable interactions between the free electron inside the cavity and the ionized molecular system. The new definition implies that ionization properties can be profoundly modified using quantum fields. Despite respecting a theoretical consistency, the methods proposed in Refs.~\onlinecite{deprince2021cavity,Equation_Liebenthal,pavosevic2022cavity,pavosevic2021polaritonic} do not account for the cavity mediated interactions between all parts of the system because the free electron is not considered explicitly. Here we develop new methodologies to include the free electron contributions.

In standard electronic structure theory, ionization potentials can be approximated using the Hartree Fock orbital energies by means of Koopmans’ theorem.\cite{helgaker2014molecular,manne1970koopmans}  
A QED extension of the theorem has not yet been developed since a consistent molecular orbital theory for polaritonic systems was not available. Recently, we solved this problem using a new \textit{ab initio} method called strong coupling QED-HF (SC-QED-HF),\cite{riso2022molecular} which allows us to formulate a QED version of Koopmans' theorem and test its accuracy.

The paper is organized as follows. In Sections \ref{sec:Long-distance} and \ref{sec:theory} we present a comprehensive theoretical framework for ionization processes in optical cavities. In particular, we provide a detailed definition of the ionization potential (IP) in the strong coupling regime and we discuss different approximations to include interactions between the ionized system and the free electron. In section \ref{sec:results} our methodologies are applied to several organic molecules to assess the relevance of the different energy contributions. The results of this section have also been compared to data from the literature. The final section contains our concluding remarks and perspectives.

\section{Long-range interactions in QED environments}\label{sec:Long-distance}
\begin{figure}
    \centering
    \includegraphics[scale=0.12]{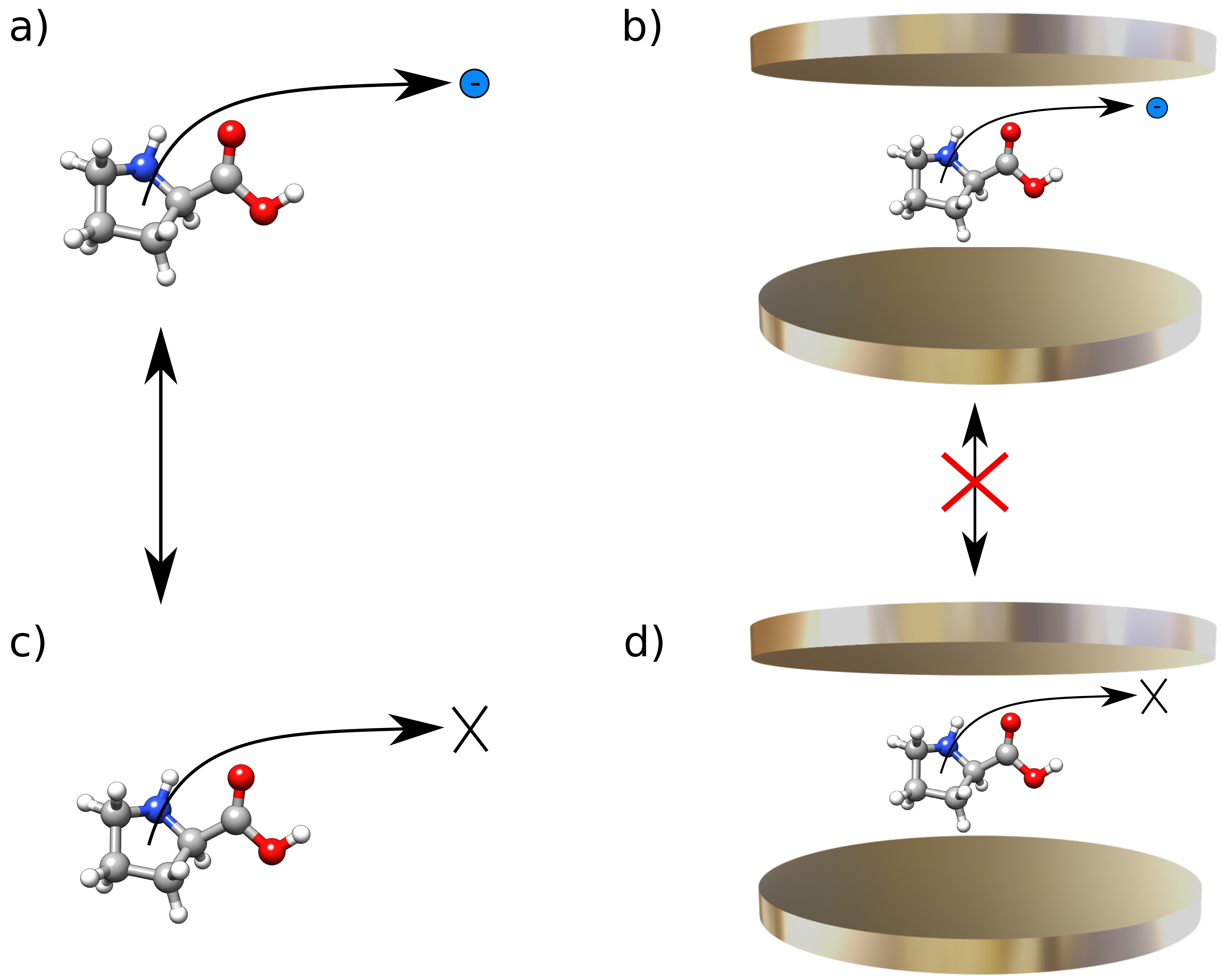}
    \caption{a) and c) Outside the cavity exciting an electron from a molecule to the continuum (free orbital) is equivalent to annihilating an electron. b) and d) This equivalence is not respected inside optical cavities. }
    \label{fig:Farò_Fig}
\end{figure}

\begin{figure}
    \centering
    \includegraphics[width=0.5\textwidth]{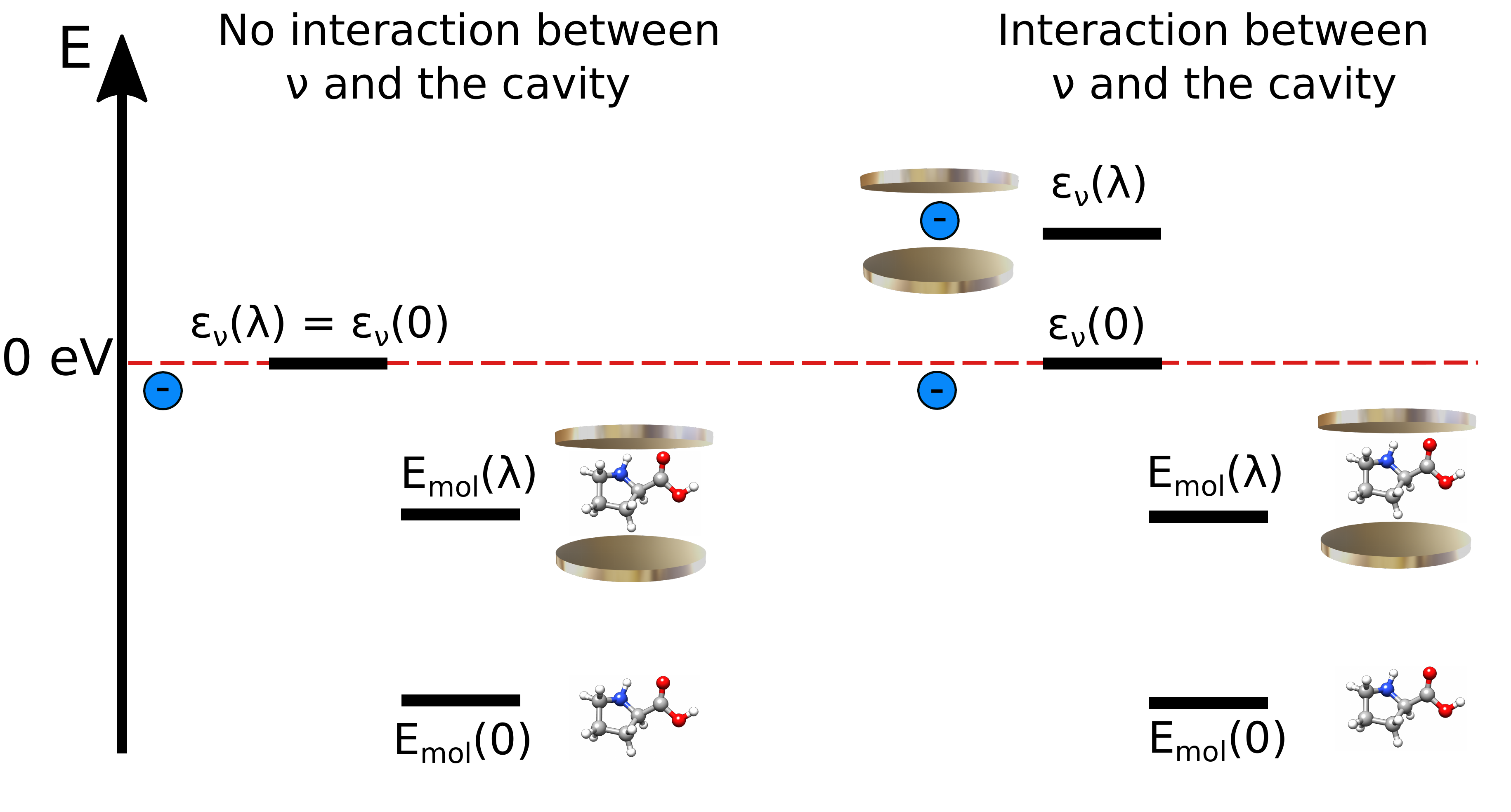}
    \caption{Pictorial representation of the cavity effect on the energy of the molecule and free electron. If the interaction between the free electron and the cavity is neglected only the energy of the molecular system increases. The energy of the free electron is instead unchanged. On the other hand, if the interaction between the free electron and the cavity is considered, the energy of both the molecule and the free electron increases.}
    \label{fig:Orbital_energies}
\end{figure}

\begin{figure}
    \centering
    \includegraphics[width=0.5\textwidth]{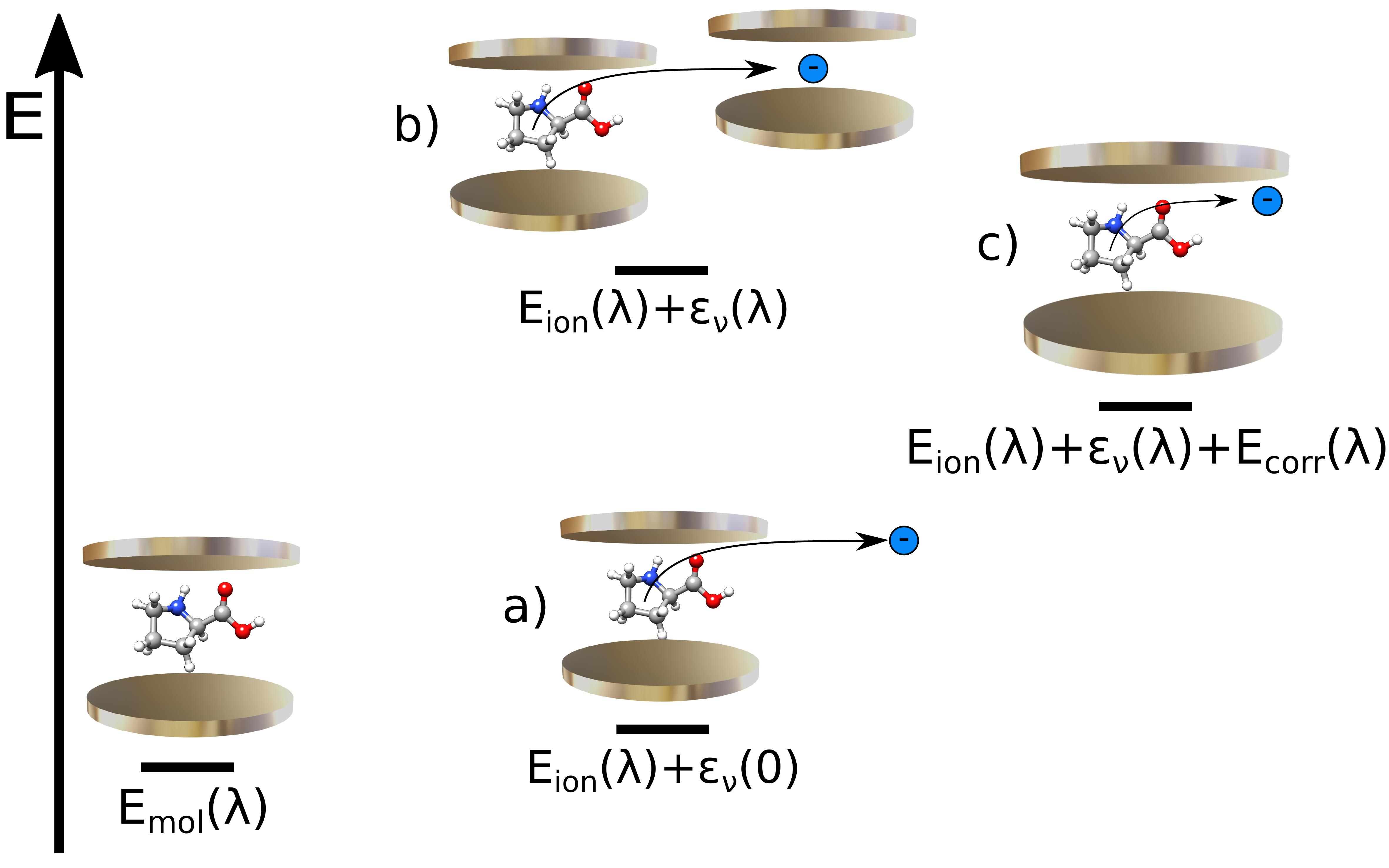}
    \caption{Different approximations to calculate IPs in an optical cavity. a) the interaction between the cavity and the free electron is neglected, b) the free electron is included, but without interaction with the molecule, c) the free electron is included also taking into account the interaction with the molecule.}
    \label{fig:Correlation}
\end{figure}

In optical cavities, the vacuum field mediates long-range interactions between molecules introducing non size-extensive effects.\cite{Intermolecular_Haugland,schafer2019modification} This means that under strong coupling conditions two molecules infinitely far apart are somehow still feeling each other. The cavity-induced non size-extensivity is a consequence of the photon coherence, meaning that all the molecules interact with the same electromagnetic field independently from their distance to each other. In simple terms, since each molecule interacts with the same cavity photons and partially changes the field shape, different molecules are indirectly coupled through the field. Because of these field-mediated long-range interactions it is reasonable to expect that the cavity photons might also introduce correlation between free electrons and molecules. Such a situation appears, for example, after ionization. In standard quantum chemical calculations, the interaction between the free electrons and the ionized molecule does not affect the IP, see Fig.\ref{fig:Farò_Fig}a and \ref{fig:Farò_Fig}c.\cite{stanton1999simple,moitra2022multi} Moreover, since the ionization potential is defined as the minimum work needed to release an electron, the outgoing free electron has zero energy. The process can therefore be described both as an excitation to a continuum orbital with zero energy or, equivalently, as the annihilation of one electron. In this work, we refer to the continuum orbital with the index $\nu$. The equivalence between the two descriptions does not hold inside optical cavities, Fig.~\ref{fig:Farò_Fig}b and \ref{fig:Farò_Fig}d, for different reasons. Firstly, the minimum free electron energy inside the cavity $\epsilon_{\nu}(\lambda)$, where $\lambda$ is the light-matter coupling strength, is larger than zero as shown in Fig.~\ref{fig:Orbital_energies} and demonstrated in Section~\ref{sec:free_electron}. Therefore, a first approximation of the ionization potential inside an optical cavity should be
\begin{equation}
IP = E_{\textrm{ion}}(\lambda)+\epsilon_{\nu}(\lambda)-E_{\textrm{mol}}(\lambda),\label{eq:Crude_approx} 
\end{equation}
where the free electron energy $\epsilon_{\nu}(\lambda)$ has been added to the standard definition of ionization potential. In Eq.~(\ref{eq:Crude_approx}),  $E_{\textrm{ion}}(\lambda)$ and $E_{\textrm{mol}}(\lambda)$ are the energies of the ionized and the non-ionized molecule inside the cavity, respectively.
However, in Eq.~(\ref{eq:Crude_approx}) we are disregarding the field coherence. This is represented in Fig.~\ref{fig:Correlation}b where the free electron and the ion are in two different cavities and therefore feel two different electromagnetic field.
An accurate treatment of the ionization problem instead requires the inclusion of field mediated correlation between the system components, $E_{\textrm{corr}}(\lambda)$, see Fig.~\ref{fig:Correlation}c. In the next section we therefore present the theoretical methodologies developed to include the contributions discussed above.
\section{Theory}\label{sec:theory}
We start this section by introducing the essential notation. 
The interaction between photons and matter will be described using the minimal coupling Hamiltonian\cite{loudon2000quantum,schafer2019modification}
\begin{align}
H =&\frac{1}{2} \sum_{i}\left(\mathbf{p}_{i}-\frac{\lambda}{\sqrt{2\omega}}\boldsymbol{\epsilon}(be^{i\mathbf{k}\cdot\mathbf{r}_{i}}+b^{\dagger}e^{-i\mathbf{k}\mathbf{r}_{i}})\right)^{2}\nonumber\\
-&\sum_{i \alpha}\frac{Z_{\alpha}}{\left|\mathbf{r}_{i}-\mathbf{r}_{\alpha}\right|}+ \frac{1}{2}\sum_{i\neq j}\frac{1}{\left|\mathbf{r}_{i}-\mathbf{r}_{j}\right|}+\omega b^{\dagger}b,
\label{eq:Singlemode}
\end{align}
where $i$ and $j$ label electrons and $\alpha$ labels nuclei with charge $Z_{\alpha}$. The cavity parameters $\omega$, $\epsilon$ and $\mathbf{k}$ are the frequency, the polarization and the wave vector, respectively. The bosonic operators $b$ and $b^{\dagger}$ annihilate and create photons. 
In this work we only consider a single cavity mode. When the field wave length is large enough compared to the molecular dimension, the dipole approximation $e^{i\mathbf{k}\cdot\mathbf{r}}\approx 1$ can be employed in Eq.~(\ref{eq:Singlemode}) leading to\cite{Light_Rokaj}
\begin{equation}
\begin{split}
H =&\frac{1}{2} \sum_{i}\left(\mathbf{p}_{i}-\frac{\lambda}{\sqrt{2\omega}}\boldsymbol{\epsilon} (b+b^{\dagger})\right)^{2} \\
-&\sum_{i \alpha}\frac{Z_{\alpha}}{\left|\mathbf{r}_{\alpha}-\mathbf{r}_{i}\right|}+\frac{1}{2}\sum_{i\neq j}\frac{1}{\left|\mathbf{r}_{i}-\mathbf{r}_{j}\right|}+\omega b^{\dagger}b.   
\label{eq:Dipole_approx}
\end{split}
\end{equation}
Using the Power-Zienau-Woolley transformation, \cite{woolley2020power,babiker1983derivation,andrews2018perspective} the length gauge form of the light matter Hamiltonian is obtained\cite{vukics2014elimination,ashida2021cavity,di2019resolution}
\begin{equation}
\begin{split}
H =& H_{e}-\lambda\sqrt{\frac{\omega}{2}}\sum_{pq}(\mathbf{d}\cdot\boldsymbol{\epsilon})_{pq}E_{pq}(b+b^{\dagger})+\omega b^{\dagger}b\\
+&\frac{\lambda^{2}}{2}\sum_{pqr}(\mathbf{d}\cdot\boldsymbol{\epsilon})_{pr}(\mathbf{d}\cdot\boldsymbol{\epsilon})_{rq}E_{pq}\\
+&\frac{\lambda^{2}}{2}\sum_{pqrs}(\mathbf{d}\cdot\boldsymbol{\epsilon})_{pq}(\mathbf{d}\cdot\boldsymbol{\epsilon})_{rs}e_{pqrs},   
\label{eq:Lenght_gauge}
\end{split}
\end{equation}
where $\mathbf{d}$ is the molecular dipole operator and we adopted the second quantization formalism for the electrons. The electronic Hamiltonian $H_{e}$ is defined as
\begin{equation}
H_{e}= \sum_{pq}h_{pq}E_{pq}+\frac{1}{2}\sum_{pqrs}g_{pqrs}e_{pqrs}, \label{eq:Standard_quantum}    
\end{equation}
where $h_{pq}$ and $g_{pqrs}$ are the one and two electron integrals, while
\begin{equation}
\begin{split}
&E_{pq}=\sum_{\sigma}a^{\dagger}_{p\sigma}a_{q\sigma} \\ 
&e_{pqrs}=E_{pq}E_{rs}-\delta_{rq}E_{ps},\label{eq:Singlet}
\end{split}
\end{equation}
with $a^{\dagger}_{p\sigma}$ and $a_{p\sigma}$ respectively creating and annihilating an electron in the orbital $p$ with spin $\sigma$.
The lowest energy eigenfunction of the Hamiltonian in Eq.~(\ref{eq:Lenght_gauge}) can be approximated using one of the \textit{ab initio} methods mentioned above. In this work, we mainly focus on two approaches:  SC-QED-HF \cite{riso2022molecular} and QED-CC\cite{Coupled_Haugland}.

\subsection{The ionization potential in QED environments}\label{sec:Ionization_definition}

In absence of light-matter interaction, ionization of a molecule can be described both as an annihilation of one electron or as an electronic excitation to the continuum (See Fig.~\ref{fig:Farò_Fig}a and \ref{fig:Farò_Fig}c).

These two approaches are equivalent. In fact, if the basis is divided into a set of localized molecular orbitals $p,q,r,s$ plus one bath orbital $\nu$ describing the continuum, the Hamiltonian $H_{e}$ takes the form
\begin{equation}
\begin{split}
H_{e} &= \sum_{pq}h_{pq}E_{pq}+\frac{1}{2}\sum_{pqrs}g_{pqrs}e_{pqrs}\\
      &+h_{\nu\nu}E_{\nu\nu}+\frac{1}{2}g_{\nu\nu\nu\nu}e_{\nu\nu\nu\nu}, 
\label{eq:Electronic}
\end{split}
\end{equation}
where all the integrals coupling the $\nu$ orbital and the molecular orbitals are equal to zero, while $h_{\nu\nu}$ and $g_{\nu\nu\nu\nu}$ are set to zero along the lines of Ref.~\onlinecite{stanton1999simple}. 
The Hamiltonian is therefore separable and the energy of the ionized system is the sum of the molecular energy plus the energy of the free electron. 
Moreover, since the free electron is assumed to have zero energy, the excitation to the $\nu$ orbital is identical to an electron annihilation. \\
The same considerations are not valid in the presence of the field, as schematized in Fig.~\ref{fig:Farò_Fig}. The minimal coupling Hamiltonian, Eq.~(\ref{eq:Dipole_approx}), is indeed equal to
\begin{align}
H =& \sum_{pq}\left(h_{pq}-\frac{\lambda(\mathbf{p}\cdot\boldsymbol{\epsilon})_{pq}}{\sqrt{2\omega}} (b+b^{\dagger})+\frac{\lambda^{2}\delta_{pq}}{4\omega}(b+b^{\dagger})^{2}\right)E_{pq}\nonumber\\  +&\left(h_{\nu\nu}-\frac{\lambda(\mathbf{p}\cdot\boldsymbol{\epsilon})_{\nu\nu}}{\sqrt{2\omega}} (b+b^{\dagger})+\frac{\lambda^{2}}{4\omega}(b+b^{\dagger})^{2}\right)E_{\nu\nu}\nonumber\\ +&\frac{1}{2}\sum_{pqrs}g_{pqrs}e_{pqrs}+\frac{1}{2}g_{\nu\nu\nu\nu}e_{\nu\nu\nu\nu}+\omega b^{\dagger}b,
\label{eq:Not-separable}
\end{align}
where, again, all the integrals coupling the $\nu$ orbital and the molecular orbitals are set equal to zero.
In Eq.~(\ref{eq:Not-separable}), the field interacts with both the free electron and the molecular system, preventing the separability of the Hamiltonian.  \\
This result is even more evident in length gauge form in Eq.~(\ref{eq:Lenght_gauge})
\begin{equation}
\begin{split}
H =& \sum_{pq}h_{pq}E_{pq}-\lambda\sqrt{\frac{\omega}{2}}\sum_{pq}(\mathbf{d}\cdot\boldsymbol{\epsilon})_{pq}E_{pq}(b+b^{\dagger})\\
+&h_{\nu\nu}E_{\nu\nu}-\lambda\sqrt{\frac{\omega}{2}}(\mathbf{d}\cdot\boldsymbol{\epsilon})_{\nu\nu}E_{\nu\nu}(b+b^{\dagger})\\
+&\frac{1}{2}\sum_{pqrs}\left(g_{pqrs}+\lambda^{2}(\mathbf{d}\cdot\boldsymbol{\epsilon})_{pq}(\mathbf{d}\cdot\boldsymbol{\epsilon})_{rs}\right)e_{pqrs}\\
+&\frac{1}{2}\left(g_{\nu\nu\nu\nu}+\lambda^{2}(\mathbf{d}\cdot\boldsymbol{\epsilon})_{\nu\nu}(\mathbf{d}\cdot\boldsymbol{\epsilon})_{\nu\nu}\right)e_{\nu\nu\nu\nu}\\
+&\frac{\lambda^{2}}{2}\sum_{pqr}(\mathbf{d}\cdot\boldsymbol{\epsilon})_{pr}(\mathbf{d}\cdot\boldsymbol{\epsilon})_{rq}E_{pq}\\
+&\frac{\lambda^{2}}{2}(\mathbf{d}\cdot\boldsymbol{\epsilon})_{\nu\nu}(\mathbf{d}\cdot\boldsymbol{\epsilon})_{\nu\nu}E_{\nu\nu}\\
+&\lambda^{2}\sum_{pq}(\mathbf{d}\cdot\boldsymbol{\epsilon})_{\nu\nu}(\mathbf{d}\cdot\boldsymbol{\epsilon})_{pq}e_{\nu\nu pq} +\omega b^{\dagger}b,
\label{eq:Lenght_gauge_2}
\end{split}
\end{equation}
where, in addition to the indirect interaction through the cavity field, there is also a purely electronic interaction term between the free electron and the ionized molecule arising from the dipole self-energy $\left(\mathbf{d}\cdot\boldsymbol{\epsilon}\right)^{2}$.\cite{Light_Rokaj} This implies that for the QED environments we have that:
\begin{itemize}
\item Annihilating an electron is not equivalent to promoting an electron to a continuum orbital;
\item The energy after the ionization will not be equal to the energy of the free electron plus the energy of the ionized molecule.
\end{itemize}
In the following sections we will develop two different particle-conserving approaches to calculate the ionization potential in the strong coupling regime using SC-QED-HF and QED-CC. We start by considering the analytic solution for one electron in the cavity and then analyze the correlation effects.

\subsection{A free electron in an optical cavity}\label{sec:free_electron}

We begin by solving the eigenvalue problem for a free electron confined in a cavity. This problem has also been investigated by Rokaj \textit{et al.} \cite{rokaj2020free} We start from the minimal coupling Hamiltonian in the dipole approximation
\begin{align}
H =&\frac{1}{2} \left(\mathbf{p}-\frac{\lambda}{\sqrt{2\omega}}\boldsymbol{\epsilon}(b+b^{\dagger})\right)^{2}+\omega b^{\dagger}b. \label{eq:minimal_free}
\end{align}
Since the momentum $\mathbf{p}$ is the only electronic operator in Eq.(\ref{eq:minimal_free}), the electronic part of the eigenfunctions are plane waves 
\begin{equation}
\begin{split}
&\ket{\psi}_{\mathbf{q}}=\sum_{n}\ket{\phi_{\mathbf{q}},n} C_{n\mathbf{q}}\\ 
&\ket{\phi_{\mathbf{q}},n}=\frac{e^{i\mathbf{q}\mathbf{r}}}{\sqrt{V}}\frac{\left(b^{\dagger}\right)^{n}}{\sqrt{n}}\ket{0_{\text{ph}}},
\label{eq:solution}
\end{split}
\end{equation}
with fixed momentum $\mathbf{q}$. In Eq.~(\ref{eq:solution}) the cavity volume is denoted by V. For a state with momentum $\mathbf{q}$, the Hamiltonian Eq.~(\ref{eq:minimal_free}) takes the form
\begin{equation}
H =\frac{q^{2}}{2}+\frac{\lambda}{\sqrt{2\omega}}\left(\mathbf{q}\cdot \boldsymbol{\epsilon}\right)\left(b+b^{\dagger}\right)+\frac{\lambda^{2}}{4\omega}\left(b+b^{\dagger}\right)^{2}+\omega b^{\dagger}b, 
\label{eq:photonic}
\end{equation}
that only depends on photonic operators.%, where $q^{2}$ is equal to the norm of the $\mathbf{q}$ vector. \\
Equation~(\ref{eq:photonic}) can be transformed into an harmonic oscillator form,  $\tilde{H}=\Omega b^{\dagger}b$ using a series of unitary rotations. We first apply a squeezed transformation $S$
\begin{equation}
S^{\dagger}bS=b\cosh{s} -b^{\dagger} \sinh{s},
\end{equation}
where
\begin{equation}
\begin{split}
&\cosh{s}= \frac{\left(\omega+\sqrt{\omega^{2}+\lambda^{2}}\right)}{2\sqrt{\omega\sqrt{\omega^{2}+\lambda^{2}}}} \\
&S = \exp \left ( \frac{1}{2} s\left(b^2 - b^{\dagger 2}\right) \right ).
\label{eq:Squeezed}
\end{split}
\end{equation}
This effectively eliminates the quadratic terms in the field.
Afterwards, we use a coherent state transformation $\hat{Z}$
\begin{equation}
Z^{\dagger}b\hat{Z}=\hat{b}-z,
\end{equation}
where
\begin{equation}
\begin{split}
&z= \frac{\left(\mathbf{q}\cdot \boldsymbol{\epsilon}\right)\lambda}{\sqrt{2\sqrt{\left(\omega^{2}+\lambda^{2}\right)^{3}}}}\\
&Z = \exp \left ( z\left(b - b^{\dagger}\right) \right ),
\end{split}
\end{equation}
to reabsorb the interaction term between the molecule and the field.
The transformed Hamiltonian $\hat{H}$ becomes
\begin{equation}
\begin{split}
& \hat{H} =\frac{q^{2}}{2}+\tilde{\omega} \;b^{\dagger}b-\frac{\left(\mathbf{q}\cdot \epsilon\right)^{2}\lambda^{2}}{2\tilde{\omega}^{2}}+\frac{1}{2}\left(\tilde{\omega}-\omega\right), 
\label{eq:Oscilla}
\end{split}    
\end{equation}
where 
\begin{equation}
\tilde{\omega}= \sqrt{\omega^{2}+\lambda^{2}}
\label{eq:Scaling}
\end{equation}
has been introduced by the squeezed transformation.
The eigenfunctions of the Hamiltonian in Eq.~(\ref{eq:Oscilla}) are the photonic occupation number states. The eigenfunctions of the Hamiltonian in Eq.~(\ref{eq:minimal_free}) then become
\begin{equation}
\ket{\psi}_{\mathbf{q}}=SZ\frac{e^{i\mathbf{q}\cdot\mathbf{r}}}{\sqrt{V}} \ket{n_{\text{ph}}} ,\label{eq:Dressed}
\end{equation}
with energy
\begin{equation}
E_{\mathrm{free}}\left(\mathbf{q},n\right) =  \frac{q^{2}}{2}+n\tilde{\omega}-\frac{\left(\mathbf{q}\cdot \boldsymbol{\epsilon}\right)^{2}\lambda^{2}}{2\tilde{\omega}^{2}}+\frac{1}{2}\left(\tilde{\omega}-\omega\right),   
\label{eq:Energy}
\end{equation}
where $n$ is the number of photons in the cavity. We point out that the free electron in Eq.~(\ref{eq:Dressed}) must respect the boundary conditions of the cavity, meaning that:
\begin{equation}
q_{z} = m\:k_{z},    \label{eq:Electron_quantization}
\end{equation}
where $m=1,2,3\;...$

\subsection{Excitation to the cavity continuum orbital}\label{sec:excitation_to_free_orbital}

When we consider ionization as an excitation to a diffuse orbital $\nu$ inside the cavity, we must account for the interaction between this continuum orbital and the cavity field. Most importantly, since the Hamiltonian is not separable (as discussed in section \ref{sec:Ionization_definition}), the free electron and the molecule still interact indirectly via the cavity field and directly via the dipole self-energy for the length gauge Hamiltonian (see Eqs.~(\ref{eq:Not-separable})  and (\ref{eq:Lenght_gauge_2}) ).
In this section we explain how we incorporate these effects. We start from the length gauge Hamiltonian in Eq.~(\ref{eq:Lenght_gauge_2}), where all the free electron interactions are mediated through the dipole operator, $(\mathbf{d}\cdot\boldsymbol{\epsilon})_{\nu\nu}$.  Notice that we do not determine an explicit expression for the continuum orbital $\nu$, nor do we use the dressed wave function found in Eq.~(\ref{eq:Dressed}). Instead, we fix the matrix element $(\mathbf{d}\cdot\boldsymbol{\epsilon})_{\nu\nu}$ to a physically reasonable value. Specifically, we require that $(\mathbf{d}\cdot\boldsymbol{\epsilon})_{\nu\nu}$ respects the following properties:
\begin{itemize}
    \item The energy of the $\nu$ orbital should be equal to the minimal energy of the free electron in the cavity. This is obtained choosing $\mathbf{q}=\mathbf{k}$ ($m=1$ in Eq.~(\ref{eq:Electron_quantization}) ) and $n=0$ in Eq.~(\ref{eq:Energy}):
    \begin{equation}
    E_{\textrm{free}}(\mathbf{k},0) =\frac{k^{2}}{2}+ \frac{1}{2}(\tilde{\omega}-\omega). \label{eq:To_ionize}    
    \end{equation}
    For consistency with the dipole approximation, we neglect $k^{2}= \omega^{2}/c^{2}$ where $c$ is the speed of light. In all the calculations presented this contribution is significantly smaller than $1\; \textrm{meV}$.
    We point out $E_{\textrm{free}}=0$ corresponds to the case where an electron has been annihilated.
    \item If the molecule is displaced by a vector $\mathbf{a}$, the dipole matrix element should change according to
\begin{equation}
(\mathbf{d}\cdot\boldsymbol{\epsilon})_{\nu\nu}\longrightarrow (\mathbf{d}\cdot\boldsymbol{\epsilon})_{\nu\nu}+\frac{Q_{tot}}{N_{e}}(\mathbf{a}\cdot\boldsymbol{\epsilon}),
\label{eq:Displace}
\end{equation}
where $N_{e}$ is the number of electrons and $Q_{tot}$ is the total charge of the system. This ensures the origin invariance of the IPs as shown in the Supplementary Material. 
\end{itemize}
From Eqs.~(\ref{eq:To_ionize}) and (\ref{eq:Displace}), the free electron dipole matrix element is defined as
\begin{equation}
(\mathbf{d}\cdot\boldsymbol{\epsilon})_{\nu\nu} = \frac{\sqrt{2E_{\mathrm{free}}}}{\lambda} -\frac{\sum_{\alpha}Z_{\alpha}r_{\alpha}}{\sum_{\alpha}Z_{\alpha}}+\frac{\sum_{\alpha}Z_{\alpha}r_{\alpha}}{N_{e}}.
\label{eq:finald}
\end{equation}
A more detailed discussion on why Eq.(\ref{eq:finald}) is a reasonable choice for the bath orbital dipole matrix element is given in the Supplementary Material.
Now that the full Hamiltonian in Eq.~(\ref{eq:Lenght_gauge_2}) has been defined, we can use one of the QED methods discussed below to calculate ionization potentials in the presence of the field. 
Although the free electron contributions are approximated, the framework presented here will still capture the cavity induced effects on the ionization process.
The two main approximations are: 
\begin{itemize}
    \item The free electron is only modeled through the dipole operator, disregarding the wave function shape;
    \item For convenience, the dipole approximation is adopted in section~\ref{sec:free_electron} for the free electron. A full minimal coupling treatment of the problem (i.e. starting from Eq.~(\ref{eq:Singlemode}) ) is needed for more accurate results as the free electron is not confined in a small region of space. 
\end{itemize}
All the approximations discussed above have the effect of underestimating the correlation effects between the components :free electron, cavity field and molecule. Despite these approximations, the interaction between the free electron and the molecule is still a sizable contribution to the energy and will be significant for the IP.

\subsection{SC-QED-HF}\label{sec:SC-QED-HF}

The SC-QED-HF wave function is defined as:
\begin{equation}
\ket{\psi}=\textrm{exp}\left(-\frac{\lambda (b-b^{\dagger})}{\sqrt{2\omega}}\left[\sum_{p}\eta_{p}\tilde{E}_{pp}+\eta_{\nu}E_{\nu\nu}\right]\right)\ket{\textrm{HF}}\otimes\ket{0},       
\label{eq:Shape}
\end{equation}
where $\tilde{E}_{pp}$ refers to orbitals that diagonalize the $(\mathbf{d}\cdot\boldsymbol{\epsilon})$ operator, $\ket{\text{HF}}$ is an electronic Slater determinant and $\ket{0}$ is the photonic vacuum.\cite{riso2022molecular} The $\eta_{p}$ parameters are orbital specific coherent state coefficients that are optimized in the ground state calculation. To fulfill the requirements presented in section \ref{sec:excitation_to_free_orbital}, the coherent state coefficient for the $\nu$ orbital must be equal to
\begin{equation}
\eta_{\nu} = -\frac{\sum_{\alpha}Z_{\alpha}r_{\alpha}}{\sum_{\alpha}Z_{\alpha}}+\frac{\sum_{\alpha}Z_{\alpha}r_{\alpha}}{N_{e}}.  
\end{equation} 
The SC-QED-HF wave function incorporates electron-photon correlation explicitly and provides origin invariant molecular orbitals. \cite{riso2022molecular} This enables us to define a consistent Koopmans' theorem for QED environments. At the same time, using the definition in Eq.~(\ref{eq:finald}), part of the correlation between the photons and the free electron can be included in the ionization treatment. In particular, we notice that if the free electron contribution is neglected ($E_{\mathrm{free}}=0$ in Eq.~(\ref{eq:finald})), the ionization potential from orbital $i$ is equal to
\begin{align}
IP=&\bra{0}\otimes\bra{HF} a^{\dagger}_{i\sigma}a_{\nu\sigma}\bar{H}a^{\dagger}_{\nu\sigma}a_{i\sigma}\ket{HF}\otimes\ket{0} -E_{SC-QED-HF}\nonumber=\\
   =&\bra{0}\otimes\bra{HF} a^{\dagger}_{i\sigma}\bar{H}a_{i\sigma}\ket{HF}\otimes\ket{0} -E_{SC-QED-HF}\nonumber=\\
=&-\epsilon_{i},
\label{eq:Ioniza}
\end{align}
where $\bar{H}$ is defined as
\begin{equation}
\bar{H}= e^{\frac{\lambda}{\sqrt{2\omega}}\sum_{p}\eta_{p}\tilde{E}_{pp}(b-b^{\dagger})} H e^{-\frac{\lambda}{\sqrt{2\omega}}\sum_{p}\eta_{p}\tilde{E}_{pp}(b-b^{\dagger})},
\label{eq:Hbar}
\end{equation}
and $\epsilon_{i}$ is the energy of the occupied orbital $i$. 
Equation~(\ref{eq:Ioniza}) shows that using SC-QED-HF, the ionization potential is equal to minus the orbital energy, in line with standard Koopmans' theorem for HF. A similar argument can be also applied to electron affinities
\begin{equation}
EA = -\epsilon_{a},  
\label{eq:Affinity}
\end{equation}
where $\epsilon_{a}$ is the energy of the unoccupied orbital $a$.
Considering the similarities, we refer to Eqs.~(\ref{eq:Ioniza}) and (\ref{eq:Affinity}) as the QED Koopmans' theorem. The recovery of Koopmans' theorem for strongly coupled systems confirms that the orbitals provided by SC-QED-HF have the same key properties as the HF orbitals.
The QED version overestimates the real ionization potential since neither the electronic
nor the photonic parts of the wave function are re-optimized
after the electron is removed from the molecule. However, we point out that the quantities in Eqs.~(\ref{eq:Ioniza}) and (\ref{eq:Affinity}) are correct to first order in the fluctuation potential. They provide a first approximation to the ionization potentials and electron affinities reported by DePrince\cite{deprince2021cavity} using equation of motion QED-CC (EOM-QED-CC).

When the free electron contribution is included, the ionization potential becomes
\begin{align}
IP=&\left\langle a^{\dagger}_{i\sigma}a_{\nu\sigma} \bar{H} a^{\dagger}_{\nu\sigma}a_{i\sigma}\right\rangle -E_{SC-QED-HF}\nonumber\\
=&E_{free}-\epsilon_{i}\label{eq:Almost_Koopmans'}\\
-&\lambda^{2}((\mathbf{d}\cdot\boldsymbol{\epsilon})_{\nu\nu}-\eta_{\nu})\sum_{p}U_{ip}((\mathbf{d}\cdot\boldsymbol{\epsilon})_{pp}-\eta_{p})U^{\dagger}_{pi},\nonumber
\label{eq:Almost_Koopmans'}
\end{align}
where $\mathbf{U}$ is the unitary transformation between the dipole and canonical bases. We notice that since
\begin{equation}
(\mathbf{d}\cdot\boldsymbol{\epsilon})_{\nu\nu}-\eta_{\nu}= \frac{\sqrt{2E_{\textrm{free}}}}{\lambda},    
\end{equation} 
Eq.~(\ref{eq:Almost_Koopmans'}) gives Eq.~(\ref{eq:Ioniza}) when $E_{\textrm{free}}=0$. Since the free electron energy is included in the calculation of the ionization potential, we refer to the level of approximation in Eq.~(\ref{eq:Almost_Koopmans'}) as Free SC-QED-HF (F-SC-QED-HF).
We point out that the first order contribution from the fluctuation potential \cite{helgaker2014molecular} in Eq.~(\ref{eq:Almost_Koopmans'}) is not equal to zero as the free electron contribution is included. This suggests that additional correlation is needed to properly describe the ionized state. Therefore, the ionization potentials obtained using F-SC-QED-HF might be less accurate than those obtained by Koopmans’ theorem for the purely
electronic case. Since the ionized state is treated as an excited state, a time dependent SC-QED-HF treatment would be more reliable. These aspects will be investigated in the future.
\subsection{EOM-QED-CC}\label{sec:QED-CC}

Accurate ionization potentials can be computed using the QED-CC approach\cite{Coupled_Haugland}. The wave function is parametrized as
\begin{equation}
\ket{\psi}= e^{T}\ket{HF}\otimes\ket{0}, 
\label{eq:CC}
\end{equation}
where $T$ is an electron-photon excitation operator
\begin{equation}
\begin{split}
T =& \sum_{ai}t^{a}_{i}E_{ai}+\frac{1}{2}\sum_{abij} t^{ab}_{ij}E_{ai}E_{bj}+...\\
+& \sum_{ai}s^{a}_{i}E_{ai}b^{\dagger}+\frac{1}{2}\sum_{aibj}s^{ab}_{ij}E_{ai}E_{bj}b^{\dagger}...\\
+& \gamma b^{\dagger}+...
\label{eq:QED-CC}
\end{split}
\end{equation}
The parameters $t^{a}_{i}, t^{ab}_{ij}$ as well as $s^{a}_{i}, s^{ab}_{ij}$ and $\gamma$ are called amplitudes where the indices \textit{i,j} and \textit{a,b} label occupied and virtual orbitals respectively. 
In the limit where all excitations are included in the $T$ operator, the parametrization in Eq.~(\ref{eq:CC}) is exact and gives the same result as QED full configuration interaction.\cite{Coupled_Haugland,Intermolecular_Haugland} The ground state wave function in Eq.~(\ref{eq:CC}) is obtained by solving the projection equations \cite{helgaker2014molecular}
\begin{equation}
\Omega_{\mu,n}=\bra{\mu,n}e^{-T}\tilde{H}e^{T}\ket{HF,0}=0,
\label{eq:project}
\end{equation}
where $\mu$ and $n$ are the electronic and photonic excitations, respectively. The ground state energy equals
\begin{equation}
E = \bra{HF,0}e^{-T}\tilde{H}e^{T}\ket{HF,0},  
\label{eq:energy}
\end{equation}
where we adopted the notation
\begin{equation}
\ket{\mu}\otimes\ket{n} = \ket{\mu,n}. 
\end{equation}
The $\tilde{H}$ operator in Eq.~(\ref{eq:project}) is the Hamiltonian in Eq.~(\ref{eq:Lenght_gauge_2}) after a coherent state rotation
\begin{equation}
\tilde{H} = e^{-z(b-b^{\dagger})}H e^{z(b-b^{\dagger})},
\end{equation}
where 
\begin{equation}
z=-\frac{\lambda }{\sqrt{2\omega}}\bra{HF}\mathbf{d}\cdot\boldsymbol{\epsilon}\ket{HF}.    
\end{equation}
The excitation energies are obtained as the eigenvalues of the Jacobian matrix $\mathbf{A}$, defined as
\begin{equation}
A_{\mu n, \rho m}=\bra{\mu,n}e^{-T}\left[H,\tau_{\rho}\left(b^{\dagger}\right)^{m}\right]e^{T}\ket{HF,0},
\end{equation}
where $\tau_{\rho}$ is an electronic excitation operator. 
Ionization potentials are readily obtained using a particle-conserving EOM-CC approach.\cite{krylov2008equation,coriani2015communication} In this case, the Jacobian matrix also includes excitations that create an electron in the continuum orbital $\nu$.\cite{stanton1999simple,moitra2022multi} Consistently to what has been done for SC-QED-HF, if the free electron energy is included in $(\mathbf{d}\cdot\boldsymbol{\epsilon})_{\nu\nu}$ ($E_{\textrm{free}}\neq 0$ in Eq.~(\ref{eq:finald})), we refer to the method as Free EOM-QED-CC (F-EOM-QED-CC).
\begin{figure*}[ht!]
    \centering
    \includegraphics[width=1.0\textwidth]{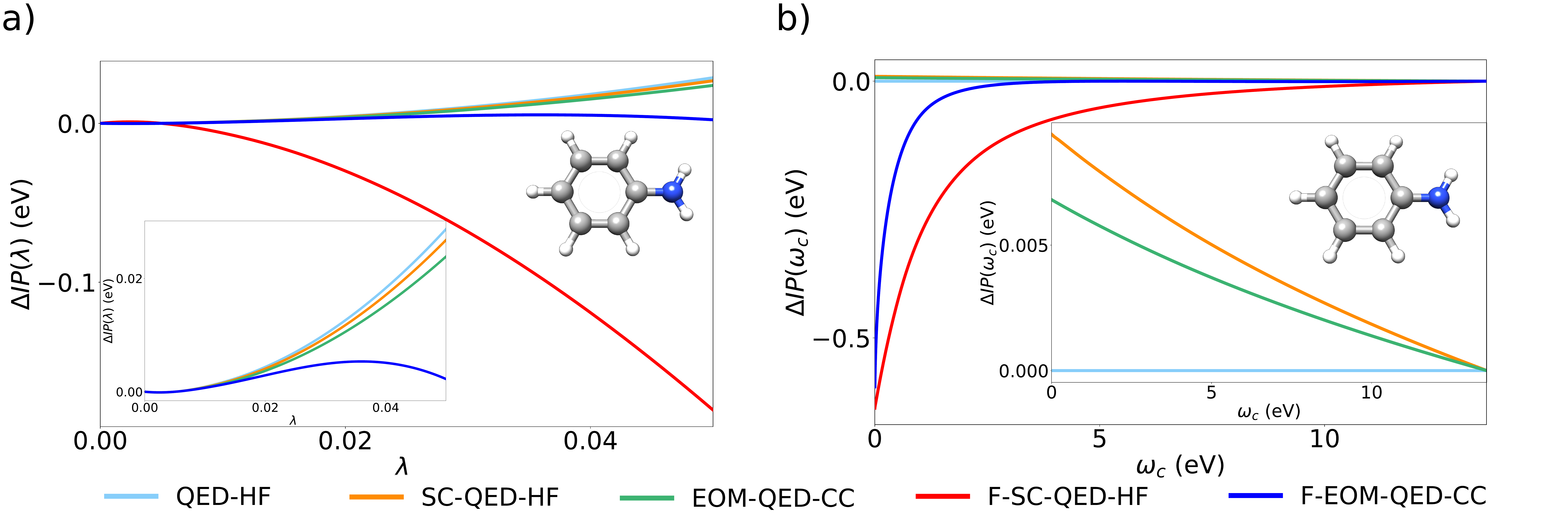}
    \caption{Dispersion of the aniline ionization potential with respect to a) the coupling $\lambda$ and b) the cavity frequency $\omega_{c}$. Calculations have been performed with different methodologies including or neglecting the free electron contribution. The cavity polarization is perpendicular to the aromatic ring.}
    \label{fig:Aniline}
\end{figure*}
\begin{table*}
\centering
    \begin{tabular}{c c  c  c }
        Method name & QED-HF & SC-QED-HF & F-SC-QED-HF \\
        \hline
        & & & \\
        Wave function & $e^{-\frac{\lambda\langle \mathbf{d}\cdot\boldsymbol{\epsilon}\rangle}{\sqrt{2\omega}}(b-b^{\dagger}}\ket{HF}\otimes\ket{0}$\hspace{5mm} & $e^{-\frac{\lambda (b-b^{\dagger})}{\sqrt{2\omega}}\left[\sum_{p}\eta_{p}\tilde{E}_{pp}+\eta_{\nu}E_{\nu\nu}\right]}\ket{\textrm{HF}}\otimes\ket{0}$ & $e^{-\frac{\lambda (b-b^{\dagger})}{\sqrt{2\omega}}\left[\sum_{p}\eta_{p}\tilde{E}_{pp}+\eta_{\nu}E_{\nu\nu}\right]}\ket{\textrm{HF}}\otimes\ket{0}$\\
        & & & \\
        Free electron energy & 0 & 0 & $0.5(\tilde{\omega}-\omega)$ \\
        & & & \\
        Ionization potential & $-\epsilon_{i}$  & $-\epsilon_{i}$ & $E_{free}-\epsilon_{i}
-\lambda^{2}((\mathbf{d}\cdot\boldsymbol{\epsilon})_{\nu\nu}-\eta_{\nu})\sum_{p}U_{ip}((\mathbf{d}\cdot\boldsymbol{\epsilon})_{pp}-\eta_{p})U^{\dagger}_{pi}$ \\
\hline\hline
    \end{tabular}
    \caption{Summary of the three different QED-HF based methods.}
    \label{tab:my_label}
\end{table*} 

\section{Results}\label{sec:results}

In this section, we use the methods presented above to compute IPs for several organic molecules (see Fig.~S1 in Supplementary Material). Here we only discuss the results for aniline as similar conclusions can be drawn from the other molecules. The data for all molecules in are reported in the Supplementary Material.
The calculations have been performed with a development version of the eT program \cite{folkestad2020t} using a cc-pVDZ basis set.\cite{dunning1989a,woon1993a,pritchard2019a,feller1996a,schuchardt2007a} The molecular geometries have been optimized using DFT-B3LYP/def2-SVP \cite{weigend1998a} basis set with the ORCA software package.\cite{The_Neese} \\
In Fig.~\ref{fig:Aniline}a, we show the dispersion of the aniline ionization potential as a function of the light-matter coupling for a fixed cavity frequency $\omega_{c}=2.0$ eV, where
\begin{equation}
  \Delta \textrm{IP}(\lambda) = \textrm{IP}(\lambda=0.0)-\textrm{IP}(\lambda).
\end{equation}
The dispersion with respect to the cavity frequency at $\lambda = 0.05$ a.u. is shown in Fig.~\ref{fig:Aniline}b, where 
\begin{equation}
\Delta \textrm{IP}(\omega_{c}) =\textrm{IP}(\omega_{c}=13.6\; \textrm{eV})-\textrm{IP}(\omega_{c}).
\end{equation}
We compare the results obtained using different levels of theory (QED-HF, SC-QED-HF, EOM-QED-CC) denoting the inclusion of the free electron contribution by an F in front of the acronym (F-SC-QED-HF and F-EOM-QED-CC). The main differences among the various QED-HF based methods are summarized in Table~\ref{tab:my_label}. The data reported in Fig.\ref{fig:Aniline} expose the crucial role played by the free electron in the ionization process. 
In particular, the comparison between the two coupled cluster based calculations (EOM-QED-CC and F-EOM-QED-CC) reveals that inclusion of the free electron leads to a major trend change in the dispersion of the ionization potential. This behaviour can be explained in terms of the qualitative concepts discussed in section \ref{sec:Long-distance}. Specifically, referring to Fig.~\ref{fig:Orbital_energies}, our results suggest that the energy of the continuum orbital $\nu$ increases more than the energy of the molecular orbitals when either the coupling increases or the frequency decreases. Moreover, since both F-SC-QED-HF and F-EOM-QED-CC include the effect of the free electron, their differences are mainly due to electron-electron and electron-photon correlation, $E_{\textrm{corr}}$ in Fig.~\ref{fig:Correlation}. As noted in section \ref{sec:SC-QED-HF}, the first order contribution from the fluctuation potential for F-SC-QED-HF is not equal to zero, therefore correlation contributions are quite large.
The SC-QED-HF (QED Koopmans' theorem) is always closer to EOM-QED-CC than standard QED-HF. In particular, while the QED Koopmans' theorem is always in qualitative agreement with EOM-QED-CC, the QED-HF method incorrectly predicts a dispersionless behaviour with respect to the frequency (see Fig.~\ref{fig:Aniline}b).\\
\begin{table*}[!ht]
\centering
\begin{tabular}{c c c c c c c c c c c c c c} 
& $\lambda$ & & QED-HF & &   SC-QED-HF & &  EOM-QED-CC  & &   $\Delta$QED-CC\cite{deprince2021cavity}         &  & F-SC-QED-HF         & &  F-EOM-QED-CC         \\
\hline
\\
\multirow{6}{*}{NaF} &0.00      & & 0.0 & &   0.0     & &    0.0   & &  0.0  & &  0.0  & &  0.0          \\
&0.01      & & -0.003  & & -0.002  & &  0.002         & &    0.00  & & -0.023& &  0.001                 \\
&0.02      & & -0.011 & & -0.008  & &  0.007            & &    0.01   & & -0.093 & &  0.002                 \\
&0.03      & & -0.025 & & -0.016  & &  0.015            & &    0.01   & & -0.203 & & -0.007                 \\
&0.04      & & -0.045 & & -0.027  & &  0.026            & &    0.03   & & -0.351 & & -0.033                 \\
&0.05      & & -0.070 & & -0.043  & &  0.039            & &    0.04   & & -0.536 & & -0.084                 \\
\hline
\\
\multirow{6}{*}{NaCl} &0.00       & & 0.0 & &  0.0     & &  0.00    & &     0.0   & &     0.0   & &   0.0        \\
&0.01       & & 0.001 & & 0.006 & &    0.003            & &  0.00    & &     0.051 & &   0.002                    \\
&0.02       & & 0.003  & & 0.013 & &    0.010            & &  0.01    & &     0.013 & &   0.006                    \\
&0.03       & & 0.007 & & 0.020 & &    0.023            & &  0.02    & &    -0.067 & &   0.006                    \\
&0.04       & & 0.011 & & 0.028 & &    0.040            & &  0.04    & &    -0.177 & &  -0.006                    \\
&0.05       & & 0.015 & & 0.037 & &    0.061            & &  0.06    & &    -0.317 & &  -0.035                    \\
\hline
\\
\multirow{6}{*}{NaBr} &0.00      & & 0.0 & &  0.0       &  &  0.00   & &   0.0    & &   0.0    & &   0.0                      \\
&0.01       & & 0.004 & & 0.004 & &   0.003              & &  0.00    & &  -0.000  & &   0.003                                \\
&0.02       & & 0.016 & & 0.016 & &   0.012              & &  0.01    & &  -0.014  & &   0.010                                \\
&0.03       & & 0.036 & & 0.030 & &   0.027              & &  0.02    & &  -0.056  & &   0.018                                \\
&0.04       & & 0.063 & & 0.053 & &   0.048              & &  0.04    & &  -0.122  & &   0.022                                \\
&0.05       & & 0.095 & & 0.079 & &   0.074              & &  0.06    & &  -0.205  & &   0.019                                \\
\hline\hline
\end{tabular}
\caption{Cavity induced variations of the ionization potential $\Delta$IP = IP(0.0)\;-\;IP($\lambda$) for sodium halides calculated at different $\lambda$ values and with different methods, including or neglecting the free electron. The variations are in eV and the cavity frequency is 2.0 eV.}
\label{tab:Table_1}
\end{table*}
In general, analogously to the standard Koopmans' theorem, SC-QED-HF seems to consistently overestimate the ionization potential compared to the corresponding coupled cluster value. The overestimation becomes even more pronounced when the free electron contribution is included because additional correlation is needed to accurately describe the ionized state.
Nonetheless, F-SC-QED-HF captures the general behaviour predicted by F-EOM-QED-CC.
Similar results can be observed for the other systems in Fig.S1. The main characteristics of the cavity induced effects can be summarized as:
\begin{itemize}
    \item The QED Koopmans' theorem is a good approximation to the ionization potential obtained using EOM-QED-CC and reproduces, at least qualitatively, the coupling and frequency dispersions.
    \item The inclusion of the free electron is needed to achieve qualitative and quantitative accuracy in the ionization potential for QED environments.
    \end{itemize}
We now compare our methods with the $\Delta$QED-CC method introduced by DePrince for sodium halides.\cite{deprince2021cavity} In Table \ref{tab:Table_1}, the cavity frequency $\omega_{c}$ is equal to $2.0$ eV, and the field polarization is along the bond axis. 
As expected, we observe a good agreement between EOM-QED-CC and $\Delta$QED-CC results. On the other hand, some differences can be observed if we compare SC-QED-HF with the results in Ref.~\onlinecite{deprince2021cavity}. Specifically, for NaF the QED Koopmans' theorem (SC-QED-HF) predicts opposite QED effects than $\Delta$QED-CC. This is the only example we have observed where SC-QED-HF does not capture the correct dispersion behaviour of the ionization potential and indicates some care should be exercised using the QED Koopmans' theorem. Nonetheless, we notice that SC-QED-HF still outperforms the results obtained using QED-HF.
As observed before, qualitative differences appear if the free electron contributions are included. In particular, the F-EOM-QED-CC data show the same trend change as the aniline results shown in Fig.~\ref{fig:Aniline}a. These differences are to be expected as the free electron is not explicitly treated in Ref.~\onlinecite{deprince2021cavity}. \\ 

\section{Conclusion}\label{sec:conclusions}

In this paper, we investigate cavity induced effects on molecular ionization processes. In particular, we provide the first consistent definition of ionization potentials and electron affinities in QED environments. In this regard, we have highlighted the crucial role played by the cavity mediated interaction between the molecule and the free electron.
Different approximations to the ionization problem have been presented using coupled cluster based methods as the reference. These approaches provide a quantification of the different effects participating in the ionization process. They also provide a benchmark methodology for ionization potentials and electron affinities. Using the recently developed SC-QED-HF theory, we formulated a QED version of Koopmans' theorem. Our work extends the investigations recently presented in Refs.~\onlinecite{deprince2021cavity,Equation_Liebenthal,pavosevic2021polaritonic,pavosevic2022cavity} on ionization processes.
The methodologies presented in this work have the same scaling as standard coupled cluster methods\cite{helgaker2014molecular} and can be applied to large number of molecules interacting with many free electrons. Studies on large systems are needed in order to investigate the relevance of collective effects on the ionization potential.
Inclusion of the complete interaction between the free electron and the molecule as well as a beyond dipole approximation treatment of the ionization process will be the subject of a future publication. This will pave the way towards providing a reference method to compute ionization potentials in the strong coupling regime without adopting a model for the free electron. Such a method is, at the moment, not available. Moreover, it will also allow us to investigate the particular configuration where the cavity is in resonance with the ionization. In that case, we expect that a Rabi splitting should be observed. However, the presented models cannot reproduce this effect because the transition dipole between ground and excited state is set to zero.
We have applied our methodologies to a set of organic and inorganic molecules and compared our results to previous calculations from the literature.\cite{deprince2021cavity,Equation_Liebenthal}
Based on our results we expect this framework to be particularly relevant to model hot electron processes. In this case, electronic excitations in nanoparticles produce free electrons that travel inside the optical cavity before being reabsorbed by a charge acceptor.\cite{shi2018enhanced} We believe the present study will provide the necessary motivation to develop experimental devices capable of measuring photoelectron spectroscopy in optical cavities. In this way, it will be possible to experimentally observe the field induced variations of ionization potentials in QED environments. 
\subsection*{Acknowledgements}
R.R.R, T.S.H and H.K. acknowledge funding from the Research Council of Norway through FRINATEK Project No. 275506. This work has received funding from the European Research Council (ERC) under the European Union’s Horizon 2020 Research and Innovation Programme (grant agreement No.  101020016). We acknowledge computing resources through UNINETT Sigma2—the National Infrastructure for High Performance Computing and Data Storage in Norway, through Project No. NN2962k.
\subsection*{Conflict of Interest}
The authors declare no conflicts of interest.
\subsection*{Data availability}
The data that support the findings of this study are available within the article and its Supplementary Material.
\bibliography{bib}
\clearpage
\end{document}